\documentclass[conference]{IEEEtran}
\IEEEoverridecommandlockouts

\usepackage{cite}
\usepackage{amsmath,amssymb,amsfonts}
\usepackage{algorithmic}
\usepackage{graphicx}
\usepackage{textcomp}
\usepackage{xcolor}
\usepackage{amsmath}
\usepackage{listings}
\usepackage{url}
\usepackage{booktabs}
\usepackage{multirow}

\def\BibTeX{{\rm B\kern-.05em{\sc i\kern-.025em b}\kern-.08em
    T\kern-.1667em\lower.7ex\hbox{E}\kern-.125emX}}
\begin{document}

\title{An Empirical Study of Reducing AV1 Decoder Complexity and Energy Consumption via Encoder Parameter Tuning
\thanks{This work was supported by the Horizon CL4 2022, EU Project Emerald, 101119800; ADAPT-SFI Research Centre, Ireland, with Grant ID 13/RC/2106\_P2; and YouTube \& Google Faculty Awards.}
}

\author{\IEEEauthorblockN{ Vibhoothi Vibhoothi$^{\dagger\star}$, Julien Zouein$^{\dagger\star}$,
Shanker Shreejith$^{\ddagger}$, Jean-Baptiste Kempf$^{\star}$, Anil Kokaram$^{\dagger}$}
\IEEEauthorblockA{
$^{\dagger}$  Sigmedia Group, $^{\ddagger}$  Reconfigurable Computing Systems Lab, \\
Department of Electronic and Electrical Engineering, \textit{Trinity College Dublin}, Dublin, Ireland \\
$^{\star}$ VideoLAN, France \\
\{vibhootv, zoueinj, shankers, anil.kokaram\}@tcd.ie
}
}

\maketitle

\begin{abstract}

The widespread adoption of advanced video codecs such as AV1 is often hindered by their high decoding complexity, posing a challenge for battery-constrained devices. 
While encoders can be configured to produce bitstreams that are decoder-friendly, estimating the decoding complexity and energy overhead for a given video is non-trivial. 
In this study, we systematically analyse the impact of disabling various coding tools and adjusting coding parameters in two AV1 encoders, libaom-av1 and SVT-AV1. 
Using system-level energy measurement tools like RAPL (Running Average Power Limit), Intel SoC Watch (integrated with VTune profiler), we quantify the resulting trade-offs between decoding complexity, energy consumption, and compression efficiency for decoding a bitstream.
Our results demonstrate that specific encoder configurations can substantially reduce decoding complexity with minimal perceptual quality degradation. 
For libaom-av1, disabling CDEF, an in-loop filter gives us a mean reduction in decoding cycles by  $\approx$10\%. For SVT-AV1, using the in-built, fast-decode=2 preset achieves a more substantial 24\% reduction in decoding cycles. These findings provide strategies for content providers to lower the energy footprint of AV1 video streaming.

\end{abstract}

\begin{IEEEkeywords}
AV1, decoding complexity, energy efficiency, rapl, socwatch.
\end{IEEEkeywords}

\section{Introduction}
The proliferation of User-Generated Content (UGC) on streaming platforms has intensified the demand for more efficient video compression. 
This growth has also led to an exponential increase in global data traffic, amplifying concerns regarding energy consumption and its correlation with greenhouse gas (GHG) emissions~\cite{2023erricsion_ghe}. In response to this, the industry has pursued solutions like the AV1 codec~\cite{2018av1paper}, which achieves 30\% bitrate savings compared to its predecessor, VP9, for the same visual quality. Along with newer codecs, new standards like the ISO's Green Metadata~\cite{ISO_IEC_23001_11_2023, fernandes2015green} have also emerged. This enables streaming systems to adapt based on decoding energy. 

The improved compression efficiency is achieved through a suite of sophisticated and computationally expensive coding tools. While this complexity is manageable in server-side encoding environments where each video is segmented into DASH chunks, this significantly increases decoder complexity on the client-side. 
This directly translates into increased energy consumption during playback, a concern for mobile and other battery-powered devices like Laptops and Tablets. This high decoding complexity directly translates to higher energy usage, which can lead to shorter device battery life and a suboptimal user experience.

To mitigate this issue, AV1 encoders like libaom-av1 and SVT-AV1 include a variety of encoder parameters which control different coding tools to produce bitstreams that are less computationally demanding to decode. 
By selectively disabling specific coding tools~\cite{kranzler2020comparative, kranzler2022optimized}, content providers can effectively reduce the decoding workload on the client-side. However, these adjustments often involve a trade-off between decoding complexity, energy consumption, and compression efficiency (i.e., video quality at a given bitrate). The precise nature of these trade-offs is not always well-documented or understood for AV1. 
Consequently, optimising the encoding process to be ``decoder complexity aware'' aids adoption of modern codecs like  AV1, especially in resource-constrained environments.

This paper analyses the impact of various encoder parameters on the energy consumption and performance of AV1 software decoding. 
We evaluate the effect of disabling individual coding tools in libaom-av1 and SVT-AV1~\cite{wu2021towardssvtav1}, including the use of fast-decode presets in SVT-AV1. 
To measure the practical decoding energy complexity, we use RAPL~\cite{2010_intelrapl, khan2018rapl} and Intel's SoC Watch~\cite{socwatch_url} tool (an extension within Intel's VTune profiler). We also present guidelines for content providers to generate energy-efficient AV1 streams with minimal impact on perceptual video quality.

\section{Background}

A typical decoding process starts with an arithmetic decoder, which reconstructs symbols from the bitstream. Subsequent stages, such as in-loop filters (e.g., Deblocking Filter), motion compensation, and various intra/inter prediction modes, all contribute significantly to the computational load and, thus, energy consumption. In a recent study, Kranzler et al.~\cite{kranzler2024comprehensive} demonstrated that dav1d, an AV1 software decoder which is heavily optimised with handwritten assembly, still requires 16.5\% more energy than libvpx-vp9, VP9 software decoder. The study also showed that VVC decoders can demand over 80\% more energy than HEVC for Random Access configurations~\cite{kranzler2020comparative}.

Several methods have been proposed to model the energy consumption of video decoding. One approach is to correlate decoding energy directly with the decoder's processing time. 
This is often linear for software decoding, making the minimisation of processing time equivalent to energy minimisation. 
In 2016, Herglotz et al.~\cite{2016_ieee_herglotz_modelinghevcenergy} demonstrated that this approach can model HEVC software decoding with less than 10\% error. However, they also observed that, depending on the duration and resolution of the video, the variance of the error can be high. The CPU measurements were conducted using RAPL and compared against the power measured via Power Meter.


In 2014, Ren et al.~\cite{ren2014energymeasure} proposed a model to estimate energy using Processor Event-based modelling for HEVC using the Valgrind profiling tool. They reported estimation errors of approximately 10\%. 
Later~\cite{herglotz2017energyprocessingevents, ramasubbu2024modelingenergyprocessevent} validated that a small subset of four key processor events (instruction reads, instruction last-level cache misses, data writes, and data write last-level cache misses) can accurately estimate software decoder processing energy for codecs like HEVC and VP9 with mean errors below 6\%. 
A limitation of this approach is the high complexity involved in deriving processor event numbers due to the instruction-level analysis required during decoding. 

Another method to measure the decoding energy is by analysing the high-level bitstream features~\cite{li2012modeling, raoufi2013energy, herglotz2015estimating} (bitrate, resolution, and time required for different coding tools). It was demonstrated to predict within a 7\% estimation error. A key advantage is that these models do not require detailed knowledge of sub-process implementations.

Encoder implementations like SVT-AV1 have introduced presets (e.g., fast-decode) that disable different tools to reduce decoding complexity at the cost of some compression efficiency in production. In 2020~\cite{kranzler2020comparative}, this type of approach is proven effective to reduce decoding complexity by disabling coding tools for VVC. 
However, a comprehensive analysis of feature-based energy estimation models specifically tailored for AV1 software decoding is not detailed in the current literature.
Given the proliferation of AV1 deployment for streaming, there is a need to optimise the energy of AV1 decoding to improve video decoding performance on consumer-grade devices. 

\section{Energy Estimation Methodology}
To assess decoding complexity, this study measures the decoding process via different methods: i) Linux perf tool, ii) RAPL, iii) Intel SoC Watch.
To isolate the computational cost of the core decoding algorithms and ensure repeatable results, all decoding tests were performed on a single thread. To simulate real-world usage, we are benchmarking on a standard x64 workstation with background processes minimised rather than on a specialised bare-metal test-bed PC (as in headless access with no background processes running).
The primary experimental method is a ``tool-off'' analysis, where different bitstreams are created with individual coding tools disabled.  Each resulting bitstream is then decoded, and its performance is compared against a baseline encode where all standard tools are enabled. This approach allows us to quantify the complexity of each tool.

\subsection{Running Average Power Limit (RAPL)}
Intel's Running Average Power Limit (RAPL~\cite{2010_intelrapl}) is a hardware feature, integrated into processor architectures since 2011 (Sandy Bridge), that allows low-overhead power monitoring. It provides access to energy consumption data through Model-Specific Registers (MSRs), which are updated approximately every millisecond for several distinct power domains in the CPU. In more modern CPUs (6th Generation), RAPL allows monitoring of the entire chip (SoC). Although RAPL relies on activity-based modelling rather than direct measurement~\cite{gough2015energy_blueprint}, its readings demonstrate high accuracy, showing a correlation coefficient of 0.99 with wall-socket power and enabling system-wide power predictions with a mean absolute percentage error as low as 1.7\%~\cite{khan2018rapl}. The performance impact of RAPL is negligible, typically below 2\%, allowing for its use in ``always-on'' monitoring scenarios. One of the key limitations is the lack of support for individual core measurements. 

\subsection{Intel SoC Watch}
Intel's SoC Watch (SocWatch) is a user‐level, kernel‐assisted telemetry tool bundled with VTune to sample RAPL-derived energy counters along with a set of platform metrics (C-states, P-states, GPU residency, memory controller activity) for every millisecond. 
Unlike direct RAPL, SoC Watch aggregates samples via its driver at a specific minimum sampling interval ($\approx$\,1\,ms) and can account for samples which can be missed due to the average. This additional accounting of the processing event can marginally increase latency. 
Thus, SoC Watch offers broader contextual visibility at the expense of resolution and minimal sampling noise, whereas direct RAPL remains preferable for fine-grained, low-overhead energy accounting of CPU-centric tasks.

\section{Experimental Setup}
For this study, we focus on the average estimated energy consumption for each decoding run. The core experimental method remains a ``tool-off'' analysis, where individual coding tools are disabled to quantify their specific impact on the overall energy footprint of the decoding process. Alongside complexity, the impact on compression efficiency and perceptual quality was evaluated using three objective metrics, PSNR, and perceptual metrics like VMAF~\cite{vmafpaper}, and UVQ~\cite{Wang_2021_CVPR_UVQ} metric.

\subsection{Dataset}
The experiments were conducted on a dataset comprising 20 video clips intended to be representative of typical online UGC. The set was composed of 10 standard-format UGC videos and 10 videos in a ``Shorts'' or vertical style. All source files were extracted from the YouTube-UGC~\cite{wang2019youtubeugc} and YouTube-SFV+~\cite{wang2024youtubeugcsfv} datasets. 
While pristine source videos are ideal for compression evaluation, using UGC source allows us to have a representative common transcoding workflow, where existing content is re-encoded before distribution.
All the videos are of 5 seconds in duration, at a resolution of 1080p, with 8-bit colour depth and frame rates of 24-30 frames per second (FPS). 
\begin{figure}
    \centering
    \includegraphics[width=\linewidth]{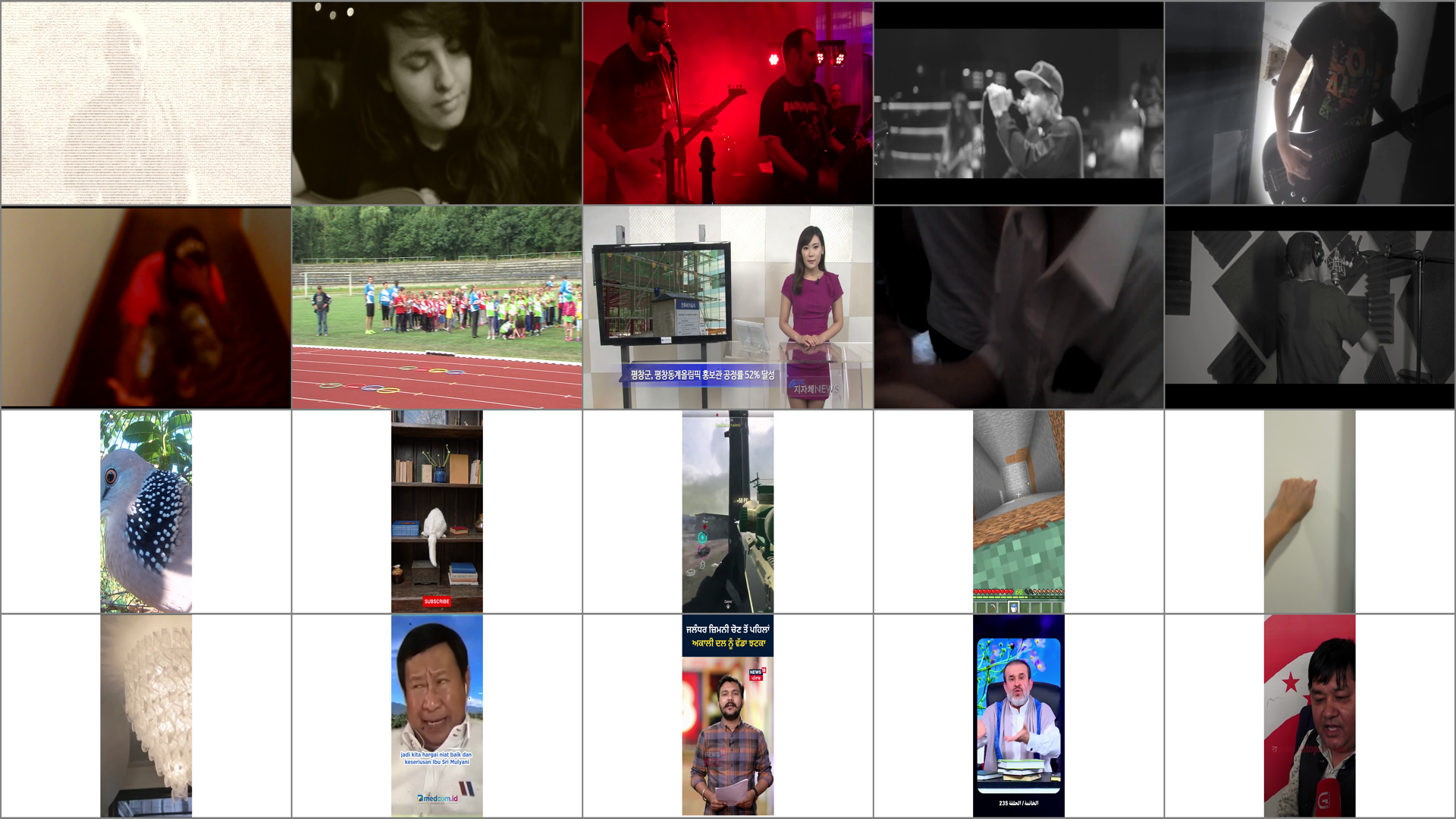}
    \caption{UGC dataset used in the study. The dataset contains a mix of Landscape and Portrait 1080p videos.}
    \label{fig:dataset-ugc}
\end{figure}

\subsection{Encoder and Decoder Setup}
We used libaom-av1 (\texttt{v3.11.0} and SVT-AV1 (\texttt{v2.3.0}) as the video encoders, dav1d (\texttt{1.5.0-0-g32cf02a}), AV1 decoder for software decoding. 
For libaom-av1, fourteen distinct flags, including four coding tools and five entropy/CDF settings (2 settings each). They were, i) Overlay Block Motion Compensation (OBMC), ii) Warped Motion, iii)  Constrained Directional Enhancement Filter (CDEF), iv) Masked Compound. The entropy coding tested was either turned off or only updated at the frame level (not updated at the block/superblock level). They were v) CDF Update modes, vi) Coefficient coding updates, vii) Transform Mode costs, viii) Motion Vector Cost statistics, ix) Displacement Vector Cost update.
For SVT-AV1, seven different flags were tested. They include i) disabling CDEF, ii) disabling Deblocking Loop Filter (dlf), iii) disabling Restoration in-loop filter, iv) disabling motion-field motion vector (mfmv), v) restricting motion vector (rmv), and vi) Two fast-decode levels. 

We have disabled tiling to reduce multi-CPU usage overhead in the analysis phase and to reduce context switches during CPU decoding.
To analyse performance under different bandwidth conditions, two target bitrates were used for encoding: 1370 Kb/s and 2316 Kb/s. In total, we have 20 videos $\times$ 15 flags $\times$ 2 Target Bitrate $\times$ 1 Tiling configuration, resulting in 600 datapoints for libaom-av1 (40 points is baseline flag), and 320 for SVT-AV1. 

The tests were conducted on x64 workstation (12th Generation Intel CPU, i7-12700K), which features a hybrid architecture of Performance-cores (P-cores) and Efficient-cores (E-cores). 
This architecture is designed to optimise the energy footprint of processes. 
We allowed the OS scheduler to manage thread placement to reflect real-world usage and observed that the decoding process was predominantly scheduled on P-cores, with only a negligible fraction of test runs (0.01\%, 8/600 encodes) utilising only E-cores, ensuring consistent and comparable performance measurements.
We chose the x86 platform for this study due to its prevalence in consumer laptops and tablets relying on software decoding for AV1, combined with the availability of robust energy profiling tools like Intel SoC Watch.


\section{Experimental Results}
This section presents the results of our evaluation, focusing on the impact of various encoder flags on decoder-side performance and active power consumption. 
Active power is calculated by subtracting the idle power of the system from the total power drawn, providing a more accurate measure of the workload of decoder. 
The results of the paper are summarised in Table~\ref{tab:obj-power-gains}. 
The dynamics of power gains over extended decoding sessions are further analysed using the heat maps in Figure~\ref{fig:decode-loop-rapl-socwatch} for both RAPL and SoC Watch.

\subsection{Analysing Overall Decoding Efficiency Gains}
The data in Table~\ref{tab:obj-power-gains} reveals a complex relationship between performance, power, and video quality. The rate control of the encoder was effective, with achieved bitrates typically within 5\% of the target for libaom-av1 and under 1\% for SVT-AV1, ensuring that bitrate variations did not significantly affect the complexity and quality comparisons.
For libaom-av1, disabling CDEF (\texttt{enable-cdef=0}) in-loop filter had the most significant impact. It reduced CPU cycles by 9.7\% and overall decode time by 11.6\%, with no loss in VMAF score and a minor drop in the UVQ metric by 0.32 points. 
For scenarios prioritising power efficiency and quality, disabling Overlay Block Motion Compensation (\texttt{enable-obmc=0}) emerged as a strong alternative, delivering the highest SoC Watch power savings (2.32\%) and RAPL gain of 1.41\% with neutral objective metrics. 

Our analysis also highlights the importance of using system-level power measurements. For instance, \texttt{mode-cost-upd-freq=3} yields a 0.88\% reduction in CPU cycles and a 1.62\% power gain at the CPU level (as measured by RAPL). Our more generalised measurements with SoC Watch reveal a contrasting 2.48\% increase in overall system power.  This discrepancy can occur because the scope of RAPL is primarily the CPU package, while SoC Watch includes other components like the memory controller. An optimisation might reduce CPU load but increase memory traffic, leading to a net power increase at the system level. This underscores that relying solely on CPU-centric data can be misleading.

For SVT-AV1, the pre-configured fast-decode preset (\texttt{fast-decode=2}) was highly effective across all tests. It reduced CPU cycles by 23.8\% and decode time by 24.6\% with a minor loss in VMAF (0.55). 
The fast-decode preset adjusts multiple encoder settings like motion search accuracy, number of reference frames in prediction hierarchy, and in-loop filter strength. Disabling individual tools like CDEF (16.2\% cycles) or the Deblocking Filter (8.6\% cycles) provided gains but were less impactful than the fast-decode preset.

\subsection{Impact of Decoding Duration on Active Power Savings}
\begin{figure*}
    \centering
    \begin{tabular}{cc}
    \includegraphics[width=0.47\linewidth]{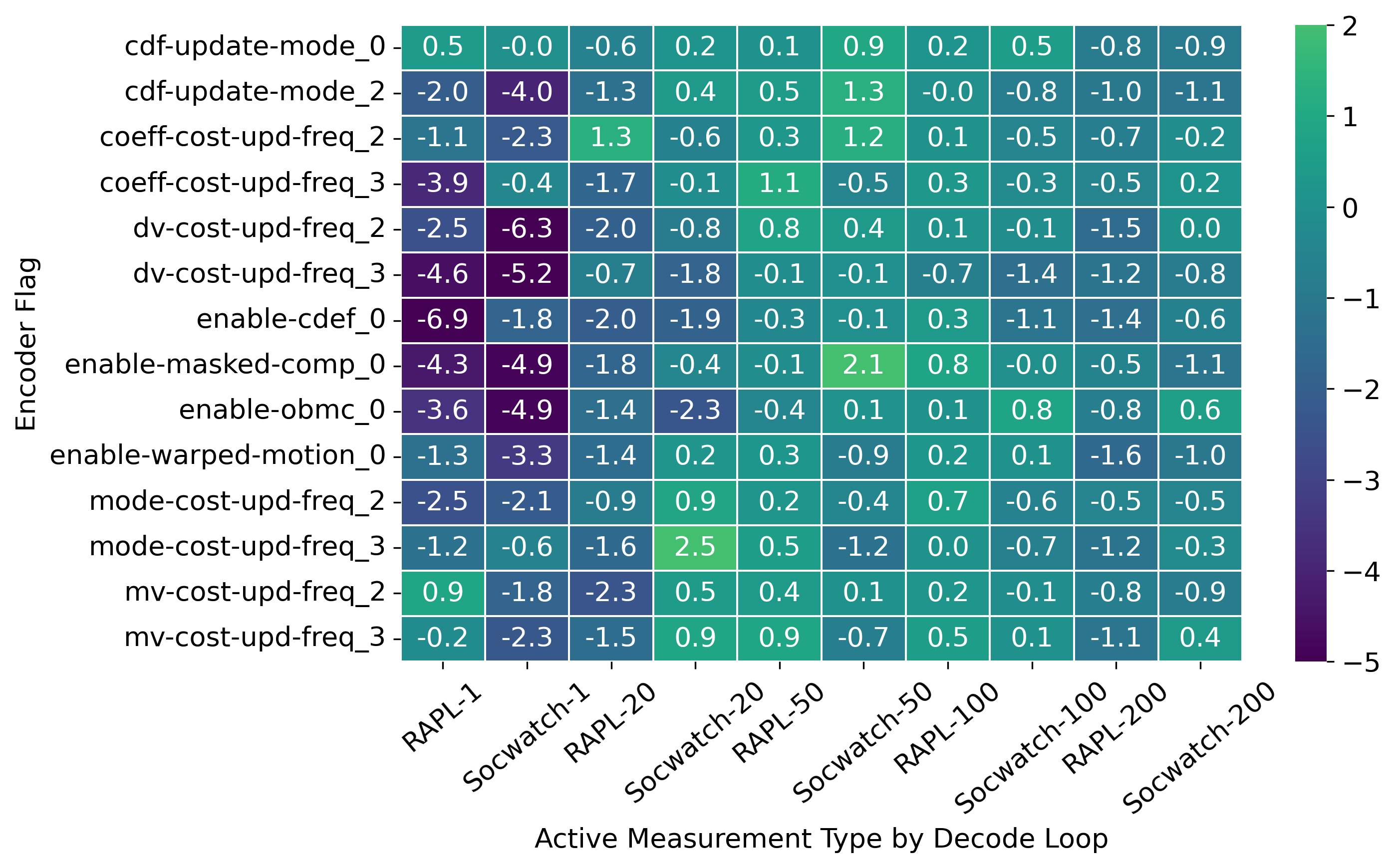} &
    \includegraphics[width=0.47\linewidth]{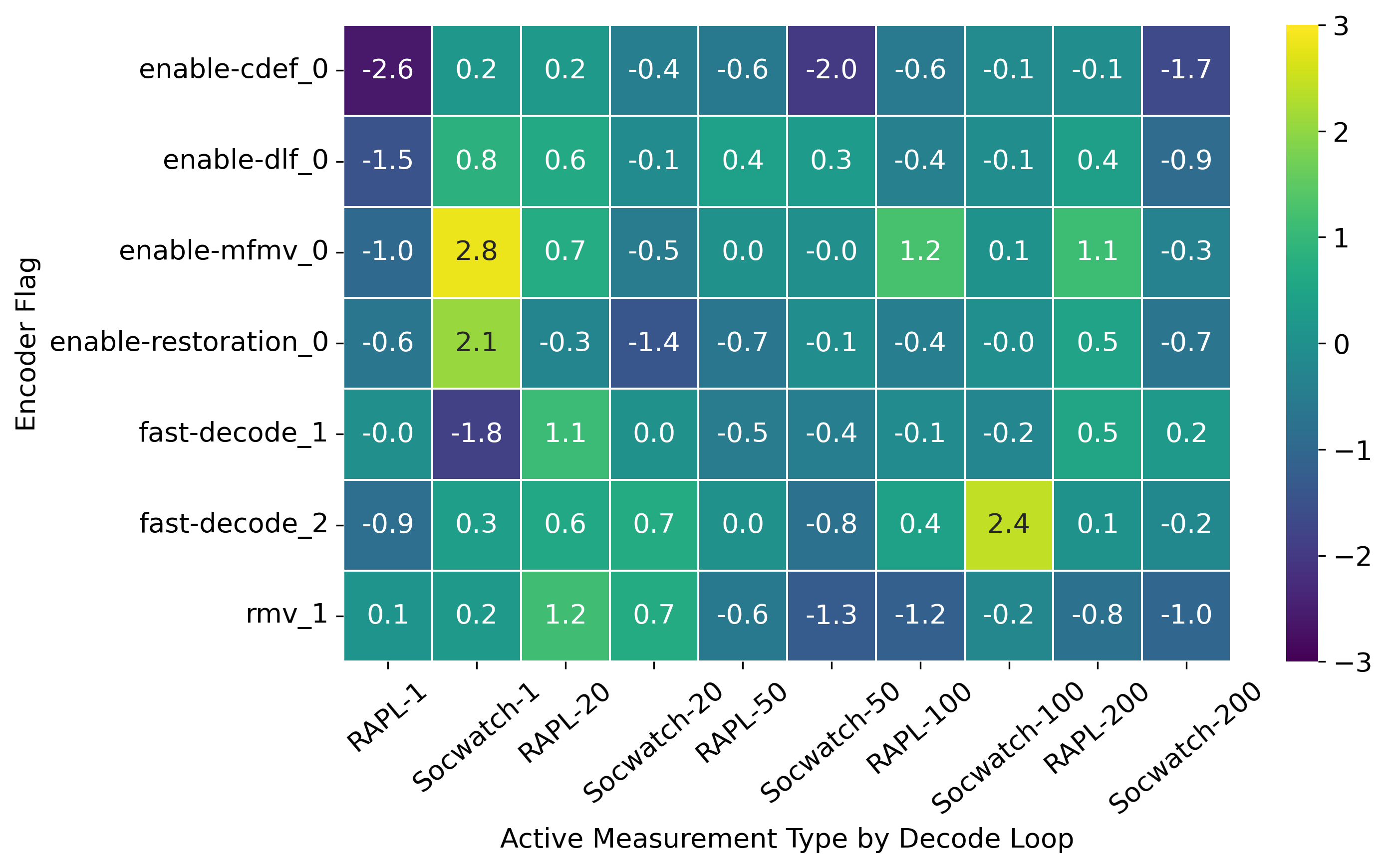} \\
    (a) libaom-av1 & (b) SVT-AV1
    \end{tabular}
    \caption{Heatmaps of Active Power Gain (\%) as a Function of Decode Duration for (a) libaom-av1 and (b) SVT-AV1. The colour of each cell represents the power saving (negative values, shown in purple, indicate greater savings) for a specific encoder flag, measured with either RAPL or SoC Watch over an increasing number of decode loops (from 1 to 200). The data illustrates that power gains are often most significant during the initial decode loops and stabilise over time.}
    \label{fig:decode-loop-rapl-socwatch}
    \vspace{-0.5em}
\end{figure*}

To analyse the stability of these gains, we measured power over extended sessions of repeated decodes. The heatmaps in Figure~\ref{fig:decode-loop-rapl-socwatch} show that the magnitude of power savings is dependent on the overall decode duration. The results in the Table~\ref{tab:obj-power-gains} is for Decode-loop of 20 (\(\sim\)30\,s). 
A key observation is that power gains are higher on the initial decode loop, with the most significant savings occurring in the first decode loop. We can also see that disabling CDEF gives 6.9\% active power gain in the first loop, while on a longer duration/pole, it results in a gain of 1.5\%. 
This initial decode load behaviour suggests that many optimisations have the highest impact during the warm-up phase of system. As the decoding process reaches a steady state, the relative power impact of these flags diminishes and stabilises. 
Comparing the two heatmaps of the two codecs reveals further insights. The flag with the highest initial gain (\texttt{enable-cdef=0}) with RAPL is different from the initial SoC Watch gain (\texttt{dv-cost-upd-freq=2}), indicating that different flags optimise power consumption differently.


\subsection{Practical Encoder Configurations Guidelines for Energy Efficient Decoding}
Based on our analysis of the most impactful configurations, we can provide targeted recommendations for different use cases:

For Maximum Performance in libaom-av1, disabling the CDEF in-loop filter gives us 12\% reduction in decode time and significant power savings, specifically 6.9\% initial SoC Watch energy reduction. 
While for SVT-AV1, the fast-decode 2 preset is most effective with $\approx$24\% reduction in decode time. This makes it well-suited for applications involving short-form video or requiring the fastest possible decode speeds. 

For Optimal Quality and Power Efficiency, disabling the OBMC in-loop filter gives 2.3\% gain in SoC Watch, to get the best overall system-level power savings with no loss in quality. This configuration is useful for scenarios looking to increase overall video playback time without reducing quality.

For Balanced Profile, refining coefficient update frequency of entropy coding (\texttt{coeff-cost-upd-freq=3}) achieves balanced gain of 0.7\% for perf cycles, 2.23\% for decode time, and 1.7\% RAPL power gain with negligible impact on quality and SoC-level power.

Finally, our findings show that system-level measurement allows us to make informed decisions, as CPU-centric metrics might not always capture the full energy picture.


\begin{table*}
\caption{Decoder Efficiency Analysis for libaom-av1 and SVT-AV1. The table compares various encoder flag configurations against a baseline encoder. Metrics include the percentage change in CPU cycles ($\Delta$Cycle) and instructions ($\Delta$Instrn), active power gains measured by RAPL and Intel SoC Watch, and change in Linux decode time ($\Delta$Time). Objective quality impact is measured by the change in VMAF, PSNR-Y, and UVQ scores. Negative values indicate a reduction in energy (gain).}
\label{tab:obj-power-gains}
\centering
\begin{tabular}{@{}lrrrrrrrrr@{}}
\toprule
\multirow{2}{*}{\textbf{Flags}} & \multirow{2}{*}{\textbf{Codec}} & \multicolumn{2}{c}{\textbf{Perf (\%)}} & \multirow{2}{*}{\textbf{\begin{tabular}[c]{@{}l@{}}RAPL \\ Gains (\%)\end{tabular}}} & \multirow{2}{*}{\textbf{\begin{tabular}[c]{@{}l@{}}SoC Watch \\ Gains (\%)\end{tabular}}} & \multirow{2}{*}{\textbf{\begin{tabular}[c]{@{}l@{}}Linux \\ $\Delta$Time (\%)\end{tabular}}} & \multicolumn{3}{c}{\textbf{Obj Metrics Loss}} \\ \cmidrule(lr){3-4} \cmidrule(l){8-10} 
 & & \textbf{$\Delta$Cycle} & \textbf{$\Delta$Instrn} &  &  &  & \textbf{$\Delta$VMAF} & \textbf{$\Delta$PSNR-Y (dB)} & \textbf{$\Delta$UVQ} \\ \midrule
cdf-update-mode=0 & libaom-av1 & -1.64 & -4.26 & -0.61 & 0.19 & -0.45 & -0.46 & -0.53 & -0.01 \\
cdf-update-mode=2 & libaom-av1 & 0.8 & -0.5 & -1.31 & 0.4 & 1.42 & -0.16 & -0.13 & 0.03 \\
coeff-cost-upd-freq=2 & libaom-av1 &  0.1 & 0.11 & 1.32 & -0.63 & -1.86 & 0.07 & 0 & 0.03 \\
coeff-cost-upd-freq=3 & libaom-av1 &  -0.71 & -0.38 & -1.7 & -0.13 & -2.23 & 0.03 & -0.03 & -0.04 \\
dv-cost-upd-freq=2 & libaom-av1 & 1.07 & 0.73 & -1.98 & -0.85 & 1.24 & 0.06 & 0.03 & 0 \\
dv-cost-upd-freq=3 & libaom-av1 & 4.13 & 0.65 & -0.73 & -1.8 & 3.59 & 0.06 & 0.03 & 0 \\
enable-cdef=0 & libaom-av1 & -9.67 & -16.78 & -2.05 & -1.89 & -11.62 & 0 & -0.04 & -0.32 \\
enable-masked-comp=0 & libaom-av1 & 1.65 & 0.51 & -1.75 & -0.37 & 1.88 & 0.08 & 0.01 & 0.13 \\
enable-obmc=0 & libaom-av1 & -0.07 & 0.4 & -1.41 & -2.32 & -0.44 & 0.06 & 0.03 & 0 \\
enable-warped-motion=0 & libaom-av1 & -1.56 & -2.02 & -1.45 & 0.22 & -1.93 & 0.05 & 0.02 & 0.08 \\
mode-cost-upd-freq=2 & libaom-av1 & 0.97 & -0.4 & -0.85 & 0.88 & 4.14 & 0.05 & 0 & -0.01 \\
mode-cost-upd-freq=3 & libaom-av1 & -0.88 & -0.88 & -1.62 & 2.48 & -2.77 & -0.24 & -0.12 & 0.05 \\
mv-cost-upd-freq=2 & libaom-av1 & 2.89 & 0.57 & -2.35 & 0.48 & 3.34 & 0.07 & 0.03 & 0.2 \\
mv-cost-upd-freq=3 & libaom-av1 & 2.78 & 0.84 & -1.46 & 0.9 & 4.57 & -0.02 & 0.01 & 0.07 \\ \midrule
enable-cdef=0 & SVT-AV1 & -16.22 & -21.9 & 0.25 & -0.45 & -16.93 & 0.18 & -0.18 & -0.53 \\
enable-dlf=0 & SVT-AV1 & -8.57 & -9.42 & 0.62 & -0.14 & -10 & 0.03 & -0.13 & -0.32 \\
enable-mfmv=0 & SVT-AV1 & -8.89 & -6.67 & 0.7 & -0.5 & -10.98 & 0.19 & 0.02 & -0.01 \\
enable-restoration=0 & SVT-AV1 & -5.05 & -1 & -0.26 & -1.4 & -7.12 & 0.13 & -0.06 & -0.12 \\
fast-decode=1 & SVT-AV1 & -15.58 & -17.27 & 1.1 & 0.01 & -15.78 & -0.55 & -0.25 & 0.12 \\
fast-decode=2 & SVT-AV1 & -23.79 & -28.74 & 0.59 & 0.66 & -24.56 & -0.55 & -0.36 & -0.08 \\
rmv=1 & SVT-AV1 & -0.57 & 1.09 & 1.16 & 0.67 & -1.35 & 0.13 & -0.15 & -0.08 \\ \bottomrule
\end{tabular}
\vspace{-0.5em}
\end{table*}

\section{Conclusion}
We found that encoder-side modification in AV1 can reduce decoding complexity by 10-24\% with minimal perceptual quality loss (UVQ, VMAF). Our analysis of libaom-av1 and SVT-AV1 shows that targeted adjustments, such as disabling the CDEF in-loop filter or using built-in presets like ``fast-decode'', can reduce decoding cycles by 10-24\%. 
We also found that system-level energy measurement with Intel SoC Watch captures the full impact of the software decoder, as simpler metrics like CPU cycles can be misleading for x64-based devices. While these findings are based on a highly optimised software decoder (dav1d), the core principle should translate to hardware decoders. Disabling computationally intensive tools at the encoder is still expected to yield power savings, though a dedicated hardware analysis would be needed to quantify the precise magnitude. Future work will analyse the impact on ARM-based mobile platforms via battery drain analysis and investigate more granular multi-parameter optimisations, including tiling behaviour.


\bibliographystyle{IEEEtran}
\bibliography{references}

\begin{thebibliography}{10}
\providecommand{\url}[1]{#1}
\csname url@samestyle\endcsname
\providecommand{\newblock}{\relax}
\providecommand{\bibinfo}[2]{#2}
\providecommand{\BIBentrySTDinterwordspacing}{\spaceskip=0pt\relax}
\providecommand{\BIBentryALTinterwordstretchfactor}{4}
\providecommand{\BIBentryALTinterwordspacing}{\spaceskip=\fontdimen2\font plus
\BIBentryALTinterwordstretchfactor\fontdimen3\font minus \fontdimen4\font\relax}
\providecommand{\BIBforeignlanguage}[2]{{%
\expandafter\ifx\csname l@#1\endcsname\relax
\typeout{** WARNING: IEEEtran.bst: No hyphenation pattern has been}%
\typeout{** loaded for the language `#1'. Using the pattern for}%
\typeout{** the default language instead.}%
\else
\language=\csname l@#1\endcsname
\fi
#2}}
\providecommand{\BIBdecl}{\relax}
\BIBdecl

\bibitem{2023erricsion_ghe}
Ericson, ``Ericson mobility report,'' 2023, {Online. Available}: \url{https: //www.ericsson.com/49dd9d/assets/local/reports-papers/mobility-report/ documents/2023/ericsson-mobility-report-june-2023.pdf}.

\bibitem{2018av1paper}
Y.~Chen, D.~Murherjee, J.~Han, A.~Grange, Y.~Xu, Z.~Liu, S.~Parker, C.~Chen, H.~Su, U.~Joshi, C.-H. Chiang, Y.~Wang, P.~Wilkins, J.~Bankoski, L.~Trudeau, N.~Egge, J.-M. Valin, T.~Davies, S.~Midtskogen, A.~Norkin, and P.~de~Rivaz, ``An overview of core coding tools in the av1 video codec,'' in \emph{2018 Picture Coding Symposium (PCS)}, 2018, pp. 41--45.

\bibitem{ISO_IEC_23001_11_2023}
\BIBentryALTinterwordspacing
ISO and IEC, ``{Information technology — MPEG systems technologies, Part 11: Energy-efficient media consumption (green metadata)},'' ISO/IEC, International Standard ISO/IEC 23001-11:2023, feb 2023. [Online]. Available: \url{https://www.iso.org/standard/83674.html}
\BIBentrySTDinterwordspacing

\bibitem{fernandes2015green}
F.~C. Fernandes, X.~Ducloux, Z.~Ma, E.~Faramarzi, P.~Gendron, and J.~Wen, ``The green metadata standard for energy-efficient video consumption,'' \emph{IEEE MultiMedia}, vol.~22, no.~1, pp. 80--87, 2015.

\bibitem{kranzler2020comparative}
{Kr{\"a}nzler, Matthias and Herglotz, Christian and Kaup, Andr{\'e}}, ``A comparative analysis of the time and energy demand of versatile video coding and high efficiency video coding reference decoders,'' in \emph{2020 IEEE 22nd International Workshop on Multimedia Signal Processing (MMSP)}.\hskip 1em plus 0.5em minus 0.4em\relax IEEE, 2020, pp. 1--6.

\bibitem{kranzler2022optimized}
M.~Kr{\"a}nzler, A.~Wieckowski, G.~Ramasubbu, B.~Bross, A.~Kaup, D.~Marpe, and C.~Herglotz, ``Optimized decoding-energy-aware encoding in practical vvc implementations,'' in \emph{2022 IEEE International Conference on Image Processing (ICIP)}.\hskip 1em plus 0.5em minus 0.4em\relax IEEE, 2022, pp. 1536--1540.

\bibitem{wu2021towardssvtav1}
P.-H. Wu, I.~Katsavounidis, Z.~Lei, D.~Ronca, H.~Tmar, O.~Abdelkafi, C.~Cheung, F.~B. Amara, and F.~Kossentini, ``Towards much better svt-av1 quality-cycles tradeoffs for vod applications,'' in \emph{Applications of Digital Image Processing XLIV}, vol. 11842.\hskip 1em plus 0.5em minus 0.4em\relax SPIE, 2021, pp. 236--256.

\bibitem{2010_intelrapl}
H.~David, E.~Gorbatov, U.~R. Hanebutte, R.~Khanna, and C.~Le, ``Rapl: Memory power estimation and capping,'' in \emph{2010 ACM/IEEE International Symposium on Low-Power Electronics and Design (ISLPED)}, 2010, pp. 189--194.

\bibitem{khan2018rapl}
K.~N. Khan, M.~Hirki, T.~Niemi, J.~K. Nurminen, and Z.~Ou, ``Rapl in action: Experiences in using rapl for power measurements,'' \emph{ACM Transactions on Modeling and Performance Evaluation of Computing Systems (TOMPECS)}, vol.~3, no.~2, pp. 1--26, 2018.

\bibitem{socwatch_url}
Intel, ``{Energy Analysis with Intel SoC Watch},'' \url{https://www.intel.com/content/www/us/en/docs/socwatch/get-started-guide/2023-1/overview.html} [Last-Access: July 2025, Online].

\bibitem{kranzler2024comprehensive}
M.~Kr{\"a}nzler, C.~Herglotz, and A.~Kaup, ``A comprehensive review of software and hardware energy efficiency of video decoders,'' in \emph{2024 Picture Coding Symposium (PCS)}.\hskip 1em plus 0.5em minus 0.4em\relax IEEE, 2024, pp. 1--5.

\bibitem{2016_ieee_herglotz_modelinghevcenergy}
C.~Herglotz, D.~Springer, M.~Reichenbach, B.~Stabernack, and A.~Kaup, ``Modeling the energy consumption of the hevc decoding process,'' \emph{IEEE Transactions on Circuits and Systems for Video Technology}, vol.~28, no.~1, pp. 217--229, 2016.

\bibitem{ren2014energymeasure}
R.~Ren, E.~Ju{\'a}rez, C.~Sanz, M.~Raulet, and F.~Pescador, ``Energy-aware decoders: A case study based on an rvc-cal specification,'' in \emph{Proceedings of the 2014 Conference on Design and Architectures for Signal and Image Processing}.\hskip 1em plus 0.5em minus 0.4em\relax IEEE, 2014, pp. 1--6.

\bibitem{herglotz2017energyprocessingevents}
C.~Herglotz and A.~Kaup, ``Video decoding energy estimation using processor events,'' in \emph{2017 IEEE International Conference on Image Processing (ICIP)}.\hskip 1em plus 0.5em minus 0.4em\relax IEEE, 2017, pp. 2493--2497.

\bibitem{ramasubbu2024modelingenergyprocessevent}
G.~Ramasubbu, A.~Kaup, and C.~Herglotz, ``Modeling the energy consumption of the hevc software encoding process using processor events,'' in \emph{2024 IEEE 26th International Workshop on Multimedia Signal Processing (MMSP)}.\hskip 1em plus 0.5em minus 0.4em\relax IEEE, 2024, pp. 1--6.

\bibitem{li2012modeling}
X.~Li, Z.~Ma, and F.~C. Fernandes, ``Modeling power consumption for video decoding on mobile platform and its application to power-rate constrained streaming,'' in \emph{2012 Visual Communications and Image Processing}.\hskip 1em plus 0.5em minus 0.4em\relax IEEE, 2012, pp. 1--6.

\bibitem{raoufi2013energy}
P.~Raoufi and J.~Peters, ``Energy-efficient wireless video streaming with h. 264 coding,'' in \emph{2013 IEEE International Conference on Multimedia and Expo Workshops (ICMEW)}.\hskip 1em plus 0.5em minus 0.4em\relax IEEE, 2013, pp. 1--6.

\bibitem{herglotz2015estimating}
C.~Herglotz and A.~Kaup, ``Estimating the hevc decoding energy using high-level video features,'' in \emph{2015 23rd European Signal Processing Conference (EUSIPCO)}.\hskip 1em plus 0.5em minus 0.4em\relax IEEE, 2015, pp. 1591--1595.

\bibitem{gough2015energy_blueprint}
C.~Gough, I.~Steiner, and W.~Saunders, \emph{Energy efficient servers: blueprints for data center optimization}.\hskip 1em plus 0.5em minus 0.4em\relax Springer Nature, 2015.

\bibitem{vmafpaper}
J.~Y. Lin, T.-J. Liu, E.~C.-H. Wu, and C.-C.~J. Kuo, ``A fusion-based video quality assessment {(FVQA)} index,'' in \emph{Signal and Information Processing Association Annual Summit and Conference (APSIPA)}, 2014.

\bibitem{Wang_2021_CVPR_UVQ}
Y.~Wang, J.~Ke, H.~Talebi, J.~G. Yim, N.~Birkbeck, B.~Adsumilli, P.~Milanfar, and F.~Yang, ``Rich features for perceptual quality assessment of ugc videos,'' in \emph{Proceedings of the IEEE/CVF Conference on Computer Vision and Pattern Recognition (CVPR)}, June 2021, pp. 13\,435--13\,444.

\bibitem{wang2019youtubeugc}
Y.~Wang, S.~Inguva, and B.~Adsumilli, ``Youtube ugc dataset for video compression research,'' in \emph{2019 IEEE 21st international workshop on multimedia signal processing (MMSP)}.\hskip 1em plus 0.5em minus 0.4em\relax IEEE, 2019, pp. 1--5.

\bibitem{wang2024youtubeugcsfv}
Y.~Wang, J.~G. Yim, N.~Birkbeck, and B.~Adsumilli, ``Youtube sfv+ hdr quality dataset,'' in \emph{2024 IEEE International Conference on Image Processing (ICIP)}.\hskip 1em plus 0.5em minus 0.4em\relax IEEE, 2024, pp. 96--102.

\end{thebibliography}

\end{document}